\documentclass[%
reprint,
superscriptaddress,
amsmath,amssymb,
aps,
prd,
]{revtex4-1}
\usepackage{graphicx}
\usepackage{dcolumn}
\usepackage{bm}
\usepackage{amsmath}
\usepackage{hyperref}
\hypersetup{hypertex=true,colorlinks=true,linkcolor=blue,anchorcolor=blue,citecolor=blue,urlcolor=blue}
\usepackage{lineno}
\usepackage{enumerate}
\usepackage{color}
\usepackage{upgreek}
\usepackage{threeparttable}
\usepackage{multirow}
\begin{document}
\preprint{APS/123-QED}

\title{Effects of Mass Diffusion on Rayleigh-Taylor Instability Under A Large Gravity}

\author{Y. Guo}
\affiliation{Key Laboratory for Laser Plasmas and Department of Physics and Astronomy, Collaborative Innovation Center of IFSA (CICIFSA), Shanghai Jiao Tong University, Shanghai 200240, People’s Republic of China}

\author{D. Wu}
\email{dwu.phys@sjtu.edu.cn}
\affiliation{Key Laboratory for Laser Plasmas and Department of Physics and Astronomy, Collaborative Innovation Center of IFSA (CICIFSA), Shanghai Jiao Tong University, Shanghai 200240, People’s Republic of China}

\author{J. Zhang}
\email{jzhang1@sjtu.edu.cn}
\affiliation{Key Laboratory for Laser Plasmas and Department of Physics and Astronomy, Collaborative Innovation Center of IFSA (CICIFSA), Shanghai Jiao Tong University, Shanghai 200240, People’s Republic of China}
\affiliation{Institute of Physics, Chinese Academy of Sciences, Beijing 100190, People’s Republic of China}%

\date{\today}

\begin{abstract}
Rayleigh-Taylor instabilities (RTI) play an important role in the evolution of inertial confinement fusion (ICF) processes, while analytical prediction of the RTI growth rate often fails to reach an agreement with the experimental and simulation results. Accurate analytical prediction of RTI growth is of great significance to the success of ICF schemes. In this paper, we study the effects of mass diffusion and exponential density distribution on RTI under a large gravity, by solving the Rayleigh equation with a linear approximation to the density distribution of the mixing layer. While both effects tend to dampen the instability growth, mass diffusion dominates the damping of perturbations of larger wavenumber and exponential density distribution dominates those of smaller wavenumber, resulting in a non-monotonicity of the density suppression factor of the instability growth rate over perturbation wavenumbers.
\end{abstract}

\maketitle

\section{Introduction}\label{sec:intro}
Rayleigh-Taylor instability (RTI) happens when the density gradient of a fluid is opposite to the pressure gradient \cite{rayleigh1882}. This condition can be interpreted as the scenario where a heavy fluid piles over a light fluid under gravity, or alternatively when a light fluid is accelerated into a heavy fluid \cite{taylor1950instability}. Driven by gravity or acceleration, a perturbation at the interface of such fluid will grow with time, leading to sinking of the heavy fluid and rising of the light fluid. It is a natural phenomena that happens everyday when someone pours water onto oil, and also has vital significance in many scientific or engineering contexts. In recent decades, Rayleigh-Taylor instability has been a subject of special interest in plasma physics, as it plays an critical role in the process of Inertial confinement fusion (ICF) \cite{atzeni2004physics} and the evolution of astronomical events like supernova remnants \cite{chevalier1992hydrodynamic}.

In typical ICF schemes, high-power lasers are shed on a fuel target to trigger intense fusion process. Common target designs consist of a light central gas and an outer dense shell, covered by a low-Z plastic ablator \cite{betti2016inertial}. This configuration is unstable to RTI both at the ablation stage and the deceleration stage, leading to mixing of fuel layers and degradation of the ICF performance. Different from traditional RTI in inviscid and immiscible neutral fluids, the fuel target during the deceleration stage is in a plasma state, possessing an extreme high temperature, high ionization degrees and will exhibit strong transport properties. Under this circumstance, electromagnetic effects \cite{evans1986influence} and plasma transport effects \cite{haines2014effects,Mikaelian1993effectofviscosity,piriz2006rayleigh,plesset1974viscous}, especially viscosity and diffusion are expected to have a strong influence on the evolution of RTI. 

Compared with the effect of viscosity, the way diffusion impact on the evolution of RTI seems to be much more complex and remain ambiguous so far. There were a lot of works dedicated to addressing the effects of diffusion \cite{Duff1962diffusion,robey2004effects} on RTI analytically, and some dispersion relations are derived with certain approximations. However, due to the miscible, time-dependent and kinetic nature of the mass diffusion behavior, a self-consistent dispersion relation is still and may be forever missing despite all these efforts. 

Under this circumstance, it is a natural result for scholars to turn to simulations for help, and kinetic simulations \cite{Yin2016Plasmakineticeffectsoninterfacialmix,Vold2021Plasmatransport,cai2021hybrid} are preferred due to its \textit{ab initio} treatment of kinetic effects just like mass diffusion and viscosity. However, results of RTI simulations often show systematic inconsistency with analytical dispersion relations, implicating defects in current theories, and leads us to improve upon the analytical modelling. Yin \cite{Yin2016Plasmakineticeffectsoninterfacialmix} and Vold \cite{Vold2021Plasmatransport} carried out a series of RTI simulations with both the kinetic code VPIC and the MHD code xRage, covering a large range of perturbation wavelength. They observed evident discrepancy between the simulation results and the theoretical dispersion relation. In order to address this problem, Keenan \cite{Keenan2023ImprovedAnalytica} carried out a comprehensive work which includes a detailed treatment of transport properties, especially for the diffusion effect, which resulted in an analytical dispersion relation much closer to the results observed in simulations. Even so, he still reported a discrepancy of the density suppression factor $\Psi$ between his analyses and the simulation results, which is related to the mass diffusion effect. Keenan attributed this discrepancy to the compressibility of the diffusion induced flow.

In this paper, we will improve the treatment of diffusion effect. Under the isothermal condition, we will consider the exponential density distribution of hydrodynamic equilibrium under the gravity. We also notice that the existence of a large gravity will accelerate the diffusion at the interface, resulting in a wider mixing layer compared to that in a isobaric diffusion. With a linear approximation to the time-dependent mass diffusion profile, we analytically solved the Rayleigh equation and gets a dispersion relation which can better explain the simulation results.

\section{General theory of RTI growth rate}
\subsection{Dispersion relation concerning transport properties}
It is a common sense that the evolution of RTI in realistic fluids or plasma is closely relevant to the transport properties. However, how the transport properties affect the linear growth rate of the instabilities remains debatable, as the effects of transport properties cannot be totally included self-consistently in the derivation of the growth rate. Moreover, determining plasma transport coefficients is a task of huge difficulty, where complex theory and lots of approximations are involved \cite{molvig2014classical}.

Over a century, many dispersion relation of viscous and diffusive RTI growth rate are established by different authors with different approaches. In ICF, most of the frequently used equations can be seen as variants of Duff's dispersion equation \cite{Duff1962diffusion}, which for the first time included the effect of diffusion:

\begin{equation}
\gamma=\sqrt{\frac{A_tgk}{\Psi}+\nu^2 k^4}-(\nu+D)k^2,    
\label{eq:dispersionduff}
\end{equation}
where $\gamma$ is the linear growth rate, $A_t$ is the Atwood number defined as $A_t=(\rho_2-\rho_1)/(\rho_2+\rho_1)$ for two fluids of uniform density $\rho_2$ and $\rho_1$, $g$ is the gravity or acceleration, $k$ is the perturbation wavenumber, $\nu$ is the viscous coefficient, $D$ is the diffusion coefficient. $\Psi$ is the density suppression factor mentioned in Sec.\ \ref{sec:intro}.

Duff derived this dispersion relation by considering three different effects: viscous effect, static diffusion effect and dynamical diffusion effect. He treated them separately to derive the dispersion equation for each effect. For the viscous effect, he adopted the equation derived by Bellman \cite{bellman1954effects}:
\begin{equation}
    \gamma=\sqrt{A_tgk+\nu^2k^4}-\nu k^2.
    \label{eq:duff1}
\end{equation}
For the static diffusion effect, Duff solved the diffusion equation for a sinusoidal interface $z=A_0 \text{sin}kx$ between two fluids. For $A_0k\ll 1$, the amplitude of the interface will damp as $dh/dt=-Dk^2h$ without the influence of gravity. As a result, Duff added this to Eq.\ \ref{eq:duff1}, where diffusion is treated as an effective viscosity:
\begin{equation}
    \gamma=\sqrt{A_tgk+\nu^2k^4}-(\nu+D) k^2.
    \label{eq:duff2}
\end{equation}
For the dynamical diffusion effect, Duff pointed out that diffusion between the two fluids will take place along with the instability growth. The diffusion process will lead to the formation of a mixing layer between two fluids, with a smooth transition of the density profile. Duff assumed this effect can be described by solving the inviscid Rayleigh equation \cite{rayleigh1882} :

\begin{equation}
    \frac{d}{dz}\left(\rho \frac{du}{dz}\right)=uk^2\left(\rho-\frac{g}{\gamma_0^2}\frac{d\rho}{dz}\right)
    \label{eq:Rayleigh}
\end{equation}
with the density distribution under the diffusion process, where $u$ is the velocity perturbation of the fluid. And as the mixing layer will always damp the instability growth, Duff substituted the square of the inviscid growth rate $\gamma_0^2$ with $A_tgk/\Psi$, where $\sqrt{A_tgk}$ is the classical RTI growth derived for two fluids of uniform density and $\Psi=A_tgk/\gamma_0^2>1$ is the density suppression factor mentioned above. In this approach, the $\Psi$ can be derived as long as the density distribution $\rho(z,t)$ is determined. Finally, Duff replaced $A_tgk$ in Eq.\ \ref{eq:duff2} by $A_tgk/\Psi$ to form Eq.\ \ref{eq:dispersionduff}.

Different from the viscous effect and the static diffusion effect, in the dynamical diffusion effect the density suppression factor $\Psi$ depends on the specific model of diffusion and the related density profile. In the preceding subsection, we will take a look on the derivation of the density profile.

\subsection{Binary diffusion and the problems}

For the dynamical diffusion effect, Duff considered the binary diffusion of two initially separated neutral fluids with uniform density $\rho_1$ and $\rho_2$ before the diffusion starts. Duff assumed the diffusion coefficient $D$ an overall constant, then the diffusion equation
\begin{equation}
    \frac{\partial \rho}{\partial t}=D\nabla^2 \rho
\end{equation}
can be easily solved with a self-similar solution:
\begin{equation}
    \rho(z,t)=\frac{1}{2}(\rho_1+\rho_2)\left[1+2A\sqrt{\pi}\int_0^\frac{z}{\epsilon}\text{exp}(-\xi^2)d\xi\right],
    \label{eq:diffusion1}
\end{equation}
where the integral in the equation is actually an error function of $z/\epsilon$, and the self-similar variable $\epsilon=\sqrt{4Dt}$ implicates the dependence of the mixing layer width on the diffusion timescale $t$. With the time-dependent diffusion density distribution Eq.\ \ref{eq:diffusion1}, Duff numerically solved $\Psi$ for different Atwood numbers. For small $k\epsilon$, Duff showed that $\Psi$ has a linear dependence on $k\epsilon$, which is similar to the conclusion of the approximate results of the variable-density situation \cite{betti1998growth}, where $\gamma=\sqrt{A_tgk/(1+A_tkL)}$, and $L$ is the minimum density gradient scale length.

However, Duff's analyses were limited to neutral fluids with uniform density distribution. For realistic plasma in ICF, two problems arise. Firstly, the density distribution on both sides of the interface is not uniform, as pressure gradient will form due to instant hydrodynamic equilibrium under the large acceleration, which will generate a density gradient in the plasma. It is obvious that the non-uniform density distribution is not self-similar in the variable of $z/\sqrt{4Dt}$, as a result solving the diffusion equation will be in principle more difficult. Secondly, the plasma binary diffusion is actually nonlinear \cite{Molvig2014NonlinnearStructure}, and modifications on the diffusion coefficient need to be made. Thirdly, under large acceleration the diffusion equation itself needs to be changed, as effects of gravity and pressure gradient on diffusion of each species need to be treated respectively while they can be considered to always cancel with each other in Duff's approach. 

Keenan noticed these problems and tried to address them with an indirect inclusion of the effect of gravity. He still assumed the fluids are always isothermal and isobaric \cite{Keenan2023ImprovedAnalytica}, but assigned an initial exponential density distribution of the lighter plasma at only around the interface to represent the effect of gravity, and the density of the heavier plasma is automatically determined to satisfy the isobaric condition. This leads to a smooth initial density distribution across the interface. Then he solved the binary diffusion equation of the light plasma with the method of Molvig \cite{Molvig2014NonlinnearStructure}. In this method, the lighter plasma is considered as the diffusing species, while the heavier plasma moves in according to the lighter plasma to satisfy the isobaric condition. The isobaric condition also allows the author to make great algebraic simplification over the diffusion equation. With the new solution of density distribution, he then followed Duff's approach to get a growth rate and explained former analytical overestimates of RTI growth compared to the simulation results \cite{Vold2021Plasmatransport}. Besides, he pointed out that an analytical growth overshoot still remains in his approach for small wavenumber and a possible reason is the effect of compressibility.

Here we basically agree with Keenan's conclusion, and a complete treatment of the density distribution of hydrodynamic equilibrium under gravity may help us improve upon this problem. To see this, we may first locate the physics missing in Keenan's approach. Firstly, the fictive initial distribution actually breaks the self-similar property of the problem, so different cutting time of the diffusion process may result in very different growth rates. It is intuitive to imagine an underestimate of the growth rate for small time scales, where the width of the actual mixing layer is much smaller compared to the fictive mixing layer. Secondly, the isothermal and isobaric approximation may be a good approximation for small gravity, but under large gravity the diffusion process must act in a very different way. 

All these problems lead us to an inspection of the diffusion process under large gravity, where the non-uniform density distribution of hydrodynamic equilibrium must be taken into consideration.

\section{Analytical modelling of the density distribution }
\subsection{Isothermal hydrodynamic equilibrium}\label{sec:model}
Over decades, most of the analyses of RTI are carried out about fluids with a uniform density distribution and only a density jump at the interface between distinct fluids. In this case, the dispersion relation of the RTI growth is simple and explicit. However, as a hydrostatic equilibrium actually requires a pressure gradient against the gravity, $\nabla P=-\rho g$, this uniform density distribution corresponds to a hydrostatic equilibrium state with a temperature gradient against the gravity, for example $n k_B\nabla T=-\rho g$ for ideal plasma, while this equilibrium only holds when heat conduction is eliminated from the system. Instead, an isothermal plasma with density gradients against the gravity may be a more natural scenario, where $k_B T \nabla n= -\rho g$, and such a distribution is often used to initialize RTI simulations \cite{Yin2016Plasmakineticeffectsoninterfacialmix,Vold2021Plasmatransport}. We should note that in realistic ICF, the instant hydrostatic equilibrium under acceleration and shock waves is established by a combination of the density gradient and temperature gradient. However, the variation of plasma transport coefficients across the area are much smaller in the isothermal case, as they roughly vary with $T^{5/2}$ but only with $\rho^{-1}$ \cite{Molvig2014NonlinnearStructure}, which allows us to study the effects of the transport properties more quantitatively.

As a result, we consider an isothermal plasma with temperature $T$ subject to a constant gravity $g$. We assume the plasma consists of two ion species, with mass $m_1< m_2$ and ionization degree $Z_1,Z_2$ separately. Also the plasma is assumed to be quasi-neutral, where electrons of the same number of charge moves according to the ions' motion, so the total pressure is $P=\sum_i(1+Z_i)n_ik_BT$ and the mass density is $\rho=\sum_i m_i n_i$, where $n_i$ is the number density of ion species $i$. At the beginning, the two ion species are separated by the interface located at $z=0$. Thus, an initial hydrostatic equilibrium can be set up by $\nabla P=k_B T\nabla n=-\rho g$, which results in 
\begin{align}
n_1(z)=&
\begin{cases}
\frac{n_0}{1+Z_1} \text{exp}[\frac{-m_1 g z}{(1+Z_1)k_B T}],& z<0\\
0,& z>0
\end{cases},\label{Eq:n1}\\
n_2(z)=&
\begin{cases}
0,& z<0\\
\frac{n_0}{1+Z_2} \text{exp}[\frac{-m_2 g z}{(1+Z_2)k_B T}],& z>0.
\end{cases}
\label{eq:distribution}
\end{align}
And the corresponding initial density distribution is
\begin{equation}
\rho(z)=
\begin{cases}
\frac{n_0 m_1}{1+Z_1} \text{exp}[\frac{-m_1 g z}{(1+Z_1)k_B T}],& z<0\\
\frac{n_0 m_2}{1+Z_2} \text{exp}[\frac{-m_2 g z}{(1+Z_2)k_B T}],& z>0.
\end{cases}
\label{eq:density}
\end{equation}
The coefficients $n_0/(1+Z_1)$ and $n_0/(1+Z_2)$ in Eq.\ \ref{Eq:n1} and Eq.\ \ref{eq:distribution} are set to maintain the continuity of pressure at the interface $z=0$, while $n_0$ is the number density of all particles at the interface. Thus, the interface is unstable to RTI as long as $m_2/(1+Z_2)>m_1/(1+Z_1)$ if we neglect the damping effects.

\begin{figure}[h]
\centering
  \begin{minipage}[t]{0.49\textwidth}
    \centering
    \includegraphics[width=0.9\textwidth]{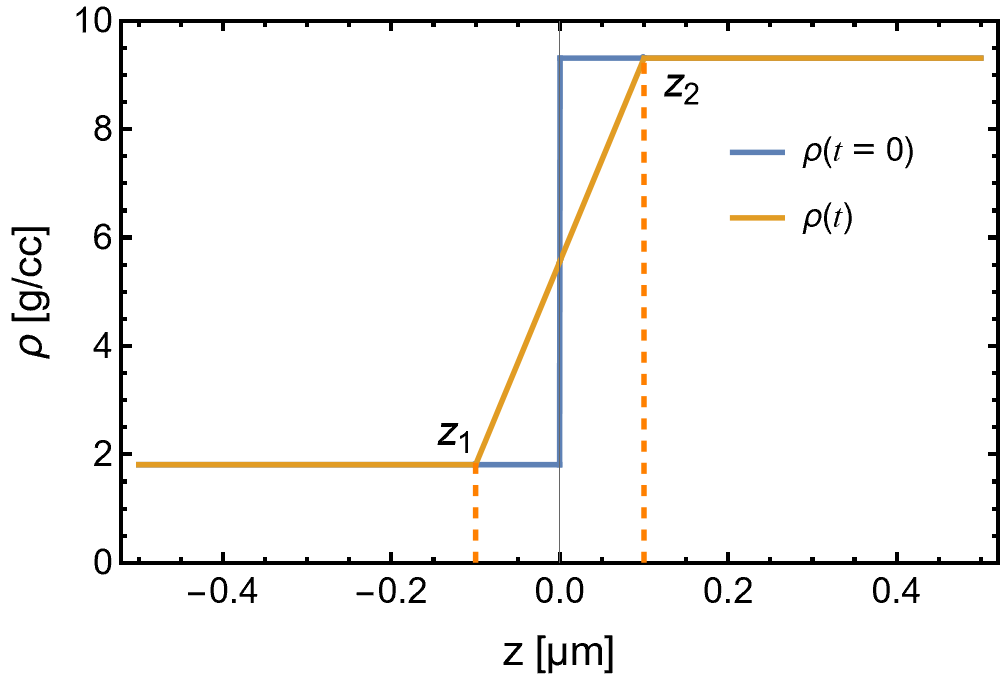}
    \caption{Density profile for the uniform density case. The choice of parameters are stated below.}
    \label{fig:diffincom}
  \end{minipage}
  \begin{minipage}[t]{0.49\textwidth}
    \centering
    \includegraphics[width=0.9\textwidth]{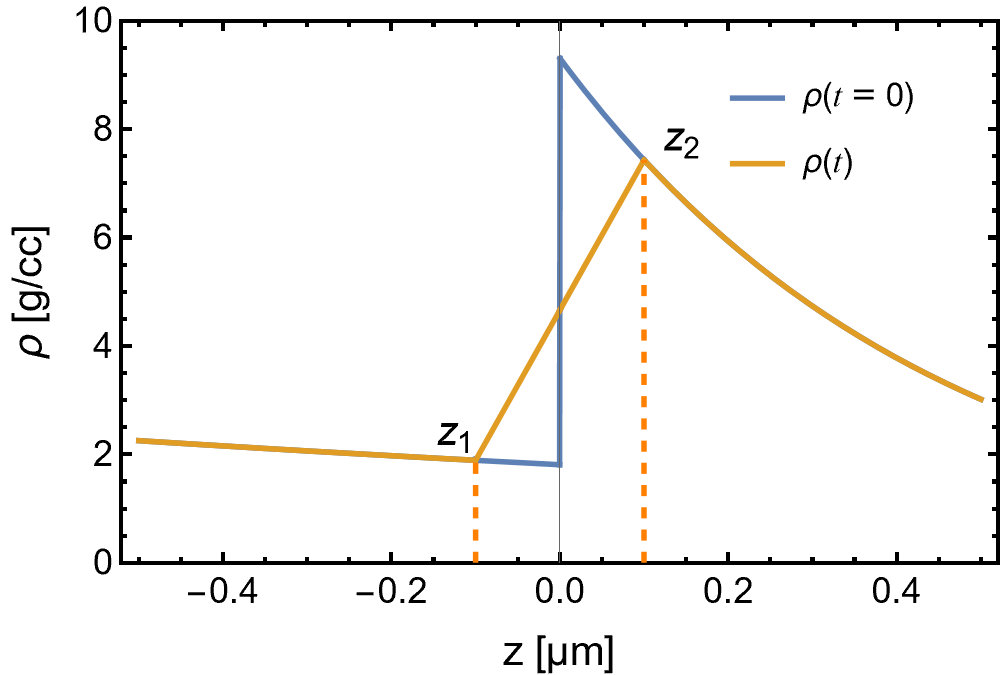}
    \caption{Density profile for the exponential density case, with the same choice of parameters as in Fig.\ \ref{fig:diffincom}.}
    \label{fig:diffcom}
  \end{minipage}
\end{figure}

As can be seen in Eq.\ \ref{eq:density}, the density distribution in our case is actually different from the uniform case Lord Rayleigh and Duff considered in their works. On the one hand, this density distribution itself will affect the solution of the Rayleigh equation Eq.\ \ref{eq:Rayleigh}, which will be discussed in Sec.\ \ref{sec:compressible}. On the other hand, when the plasma begins to diffuse, the morphology around the interface will be very different. As the mixing layer broadens, the effective Atwood number will decrease rapidly if we take the exponential distribution into consideration, as seen in Fig.\ \ref{fig:diffincom} and Fig.\ \ref{fig:diffcom}.
In these two figures, we compare the initial and mixed density profile of both the uniform density case and the exponential density case. The linear distribution of the mixing layer is not strict and only used for rough illustration. The initial density at either side of the interface is the same for both cases, while it extends to infinity for the uniform case and decays in an exponential form for the exponential case. If we assume the mixing layer has the same width for both cases, the density contrast at both ends of the mixing layer of the exponential case will be much smaller than the uniform case, which must lead to a smaller effective Atwood number. Moreover, the binary diffusion process in the exponential case will be actually faster as we take the effect of gravity into consideration, which will strengthen the effect of diffusion and dampens the instability growth.

For the sake of comparison with Keenan's conclusions \cite{Keenan2023ImprovedAnalytica}, in preceding sections we will adopt the parameters settings in simulations by Yin and Vold \cite{Yin2016Plasmakineticeffectsoninterfacialmix, Vold2021Plasmatransport} when illustrations are needed, as in Fig.\ \ref{fig:diffincom} and Fig.\ \ref{fig:diffcom}. The parameters are $m_1=2m_p,\ m_2=36m_p,\ Z_1=1,\ Z_2=6,\ g=2.1\times10^{19}\ \text{cm/s}^2,\ T=5\ \text{keV},\ n_0=1.08\times10^{24}\ \text{cm}^{-3}$. This results in an Atwood number $A_t=0.67$ defined at the initial interface. Moreover, Coulomb logarithm is assumed a constant $\ln \Lambda=100$, which allows us to determine the diffusion coefficient $D$. We must note that, the gravity $g$ is about 100 times larger than the realistic acceleration in ICF deceleration stage, and the temperature $T$ is a few times higher than that in the deceleration stage, while the Coulomb logarithm are also exaggerated by about 5 times. The authors tailored these parameters in order to carry out PIC simulations in an acceptable computation time, where a larger gravity $g$ leads to faster instability growth and a larger temperature $T$ leads to larger Debye's length of the plasma, which allows grids with lower resolution. The large Coulomb logarithm  $\ln \Lambda$ reduces the transport coefficients of plasma at the high temperature, otherwise the instability perturbation will be totally damped by diffusion.

\subsection{Diffusion process in the exponential density case}

Now that we have located the problem, it is a natural thought to solve the diffusion equation of the exponential case. However, this task presents much larger difficulty than in the isobaric case. To see that, let's take a glance at Keenan's approach to the diffusion problem. Following Molvig's treatment \cite{Molvig2014NonlinnearStructure}, he expressed the light ion mass flux as
\begin{widetext}
\begin{equation}
    m_1\Gamma_1 =-y_2 n_1 m_1 D_{12}\bigg[\alpha_{11}\bigg(\nabla \ln x_1+\frac{x_1-y_1}{n_1T_1}\nabla P_1    +\frac{z_i-y_i}{n_i T_i}\nabla P_e+ \frac{Z_\text{eff}-1}{Z_\text{eff}}\frac{\alpha_T}{T_1}\nabla T_e\bigg)+\frac{3}{2}\alpha_{12}\nabla \ln T_1\bigg],
\end{equation}
where the variables $x_j,y_j,z_j$ denote atom, mass and charge fractions respectively with $j=1$ for light ion and $j=2$ for heavy ion. Although this equation looks complex, under isothermal and isobaric conditions it reduces to
\begin{equation}
    m_1\Gamma_1=-y_2n_1 m_1 D_{12}\left[ \alpha_{11}(\Delta_2)\left(\nabla \ln x_1 +\frac{z_1-x_1}{n_1T_1}\nabla P_e\right) \right].
    \label{eq:flux}
\end{equation}
\end{widetext}
And after some more algebra the diffusion equation $\partial \rho_1/\partial t+\nabla\cdot m_1 \Gamma_1=0$ finally becomes
\begin{equation}
    \frac{\partial a}{\partial t}=\frac{\partial}{\partial x}D_0[d_c(a)+d_p(a)]\frac{\partial a}{\partial x},
    \label{eq:diffuse}
\end{equation}
where $a=\rho_1/\rho$ and $d_c(a),d_p(a)$ are functions of $a$ and other known parameters only. Thus, this equation can be easily solved, and when we derive the distribution of the light ion, the distribution of the heavy ion can be immediately determined under the isobaric condition. 

However, there are two factors preventing us from following the same approach in the exponential case. Firstly, the isobaric condition $P=\text{const}.$ results in 
$(1+Z_1)n_1k_BT+(1+Z_2)n_2k_BT=\text{const}.\ ,$
which under isothermal condition allows us to determine $n_2(z_0)$ immediately if we know $n_1(z_0)$ at $z_0$, as the total pressure is always equivalent to the initial pressure all across the area. This leads to a large amount of algebraic simplification that leads to Eq.\ \ref{eq:flux} and Eq.\ \ref{eq:diffuse}, where $x_j,y_j,z_j$ and $P_1,P_e$ are all completely determined by $n_1$ at the same location.  However, in the exponential case, the isobaric condition needs to be replaced by $\nabla P=-\rho g$, which is $(1+Z_1)k_BT\nabla n_1+(1+Z_2)k_BT\nabla n_2=-(m_1n_1+m_2n_2)g.$
It is straightforward to see that merely knowing $n_1(z_0)$ at $z_0$ will not help us know $n_2(z_0)$. In principle, we can't solve $n_2(z)$ unless we know the overall distribution of $n_1(z)$. This prevents us to perform the same simplification as in the isobaric case.

The other factor originates from the ``gas-metal" model of the diffusion process, where the light ion is considered as the diffusion species and the heavy ion only moves accordingly. This maybe a good approximation under the isobaric condition, but as we take gravity into consideration, the heavy ion may play a more important role in the diffusion process. 

After all, these difficulties turn the diffusion problem of the exponential case to a complex one, and to our knowledge no self-consistent analytical solution has been given to this specific problem. Even if it exists, it might not help much for us to determine the growth rate analytically. One may turn to PIC simulations for immediate help, and at least this will help us to validate some approximations.

\subsection{Linear approximation of the mixing layer}

Here we present a simple approximation to the diffusion process, where the density profile of the mixing layer is assumed linear, which may be qualified for our suppose at least for the early stage of diffusion. Also, this approximation allows us to study RTI in a more analytical way, as the solution of linear density distribution is more accessible than other complex profiles \cite{cherfils2000analytic}. In general, we express the overall time-dependent density distribution as
\begin{widetext}
\begin{equation}
    \rho(z)=\begin{cases}
    \rho_{10}\text{exp}[-\frac{m_1gz}{(1+Z_1)k_B T}],& \text{if } z < z_1(t), \\
    \rho_1+\frac{\rho_2-\rho_1}{z_2-z_1}(z-z_1),& \text{if } z_1(t)\leq z \leq z_2(t),\\
    \rho_{20}\text{exp}[-\frac{m_2gz}{(1+Z_2)k_B T}],& \text{if } z > z_2(t),
    \end{cases}
    \label{eq:profile}
\end{equation}
\end{widetext}
where $\rho_i=\rho_{i0}\text{exp}[-m_igz_i/((1+Z_i)k_B T)]$, while $z_i(t)$ needs to be determined. From $z_1(t)$ to $z_2(t)$ is the mixing layer with a linear density distribution and outside the layer the exponential distribution of each species remains as it was.

Now in order to solve the growth equation, we only need to determine the time-dependent boundaries $z_i(t)$, where some approximations are needed. In general, we choose to describe $z_i(t)$ as the combination of two different parts, where
\begin{align}
    z_1(t)&=-\sqrt{4Dt}+u_{s2}t, \label{eq:z1}\\
        z_2(t)&=\sqrt{4Dt}+u_{s1}t.
        \label{eq:z2}
\end{align}
The first part $\sqrt{4Dt}$ in Eq.\ \ref{eq:z1} and Eq.\ \ref{eq:z2} describes the concentration driven diffusion of the ions, which corresponds to the width of the mixing layer in the isobaric diffusion. We note that in isobaric case, the span of the mixing layer actually depends on the properties of the ion species and is different on each side, but in general the width of the layer can usually be evaluated by a few times of $\sqrt{Dt}$  \cite{Molvig2014NonlinnearStructure}, and such approximation is often used to evaluate the width of the mixing layer in order to calculate the RTI growth \cite{Yin2016Plasmakineticeffectsoninterfacialmix}.

The second part $u_{si}t$ describes the gravitational sedimentation of the different ions, where the heavy ions sediment and the light ions rise up. Note that $z_1$ is related to $u_{s2}t$, as the lower boundary of the mixing region is decided by the front of the heavy ion 2. This process happens naturally as we take gravity into consideration, while it was completely neglected in the isobaric diffusion treatment. This process is driven by the same mechanism as RTI, but it can be activated without an initial perturbation, as the kinetic diffusion will always provide the exchange of different ions.

According to Beznogov \cite{beznogov2013diffusion}, the sedimentation velocity of the ions in a binary mixture can be evaluated by

\begin{equation}
    u_{s2}=\frac{\rho_1 nD}{\rho n_e k_B T}Z_1Z_2m_p\left(\frac{A_1}{Z_1}-\frac{A_2}{Z_2}\right)g,
\end{equation}
where $A_i$ is the mass number of the ion $i$. For a heavy ion in the unmixed region, the sedimentation velocity vanishes as $\rho_1=0$. And for a heavy ion in the diffusion front, $\rho \approx \rho_1$ and $n_e\approx Z_1 n_1$, so the sedimentation velocity can be evaluated as constant:

\begin{equation}
    u_{s2}=-\frac{D}{k_B T}Z_2m_p\left(\frac{A_2}{Z_2}-\frac{A_1}{Z_1}\right)g.
\end{equation}
The same analysis leads to
\begin{equation}
    u_{s1}=\frac{D}{k_B T}Z_1m_p \left(\frac{A_2}{Z_2}-\frac{A_1}{Z_1}\right)g.
\end{equation}
Note that $u_{s1}$ and $u_{s2}$ are in the opposite directions.

\begin{figure}[htbp]
    \centering
    \includegraphics[width=0.9\linewidth]{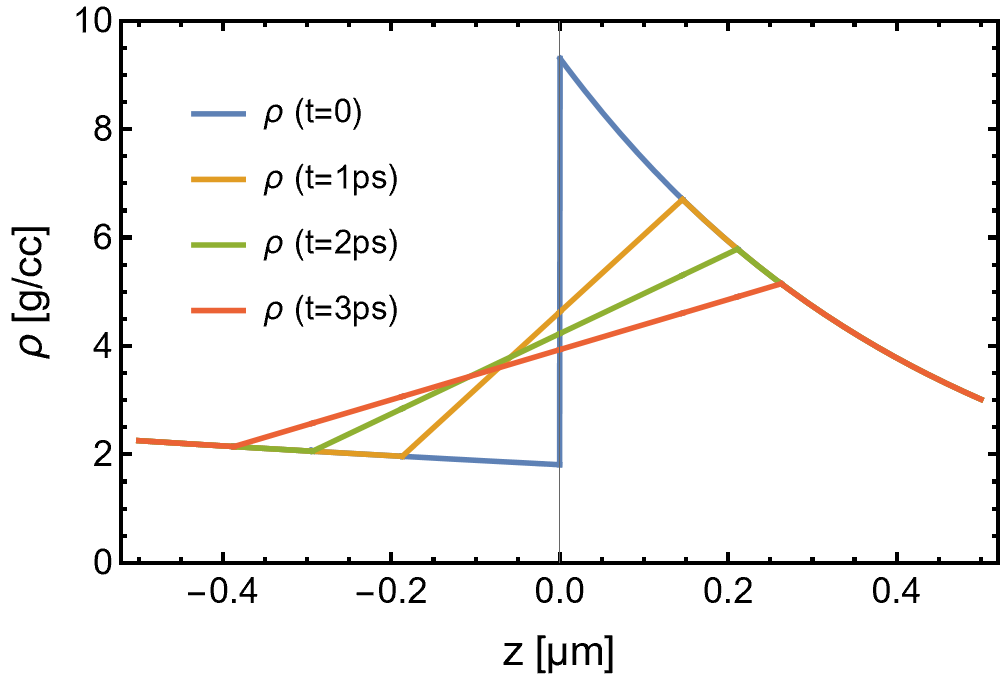}
    \caption{The time-dependent density profile determined by Eq.\ \ref{eq:profile} at different diffusion timescales.}
    \label{fig:profile}
\end{figure}

Moreover, we assume the diffusion coefficient constant, which is evaluated by \cite{Molvig2014NonlinnearStructure}
\begin{equation}
    D_0=27.8\frac{4}{\ln \Lambda}\frac{A_1^{1/2}T^{5/2}}{\rho_0}\frac{Z_2+1}{Z_2^2}\ {\mu} \text{m}^2\text{/ns}
\end{equation}
for T in keV and $\rho_0$ in g/cc. We choose $\rho_0$ as the initial density of the light ion at the interface. Now all variables in Eq.\ \ref{eq:z1} and Eq.\ \ref{eq:z2} can be determined. The resulting density distribution at different times are shown in Fig.\ \ref{fig:profile}.

 To validate our approximation, 1-D binary diffusion simulations are carried out with a full kinetic PIC code LAPINS. The initial distribution of ions are set as Eq.\ \ref{eq:distribution}, and the parameters are $m_1=2m_p,\ m_2=36m_p,\ Z_1=1,\ Z_2=6,\ g=2.1\times10^{19}\ \text{cm/s}^2,\ T=5\ \text{keV},\ n_0=1.08\times10^{24}\ \text{cm}^{-3}$ and $\ln\Lambda=100$ as in Sec.\ \ref{sec:model}. Corresponding amount of electrons are added to maintain the electric neutrality. The length of the simulation area is 4.2$\mu$m and separated into 3360 grids, and both boundaries of the simulation domain are set to be particle thermal, which means particles escaping the boundary will be re-injected from a Maxwellian distribution of the local temperature. For the fields, the boundaries are set to be absorbing.

\begin{figure}[h]
\centering
  \begin{minipage}[t]{0.49\textwidth}
    \centering
    \includegraphics[width=0.9\textwidth]{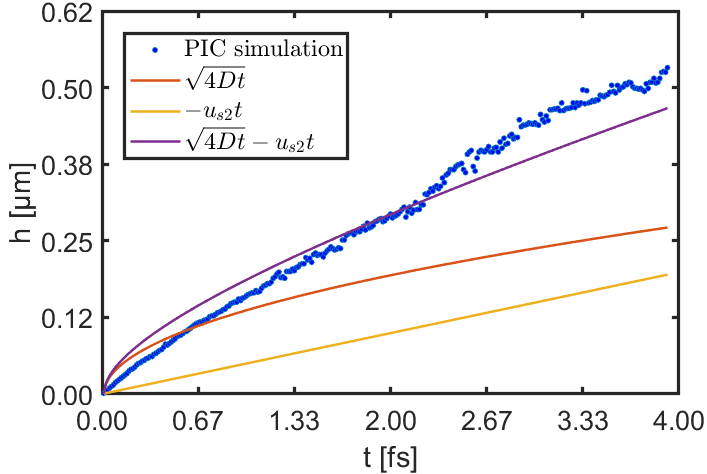}
    \caption{The diffusion height of C$^{6+}$ ion. The blue dots is measured from the simulation, the red and orange curve corresponds to the first and second part of Eq.\ \ref{eq:z1} respectively and the purple curve is the total expression of Eq.\ \ref{eq:z1}.}
    \label{fig:Cfront}
  \end{minipage}
  \begin{minipage}[t]{0.49\textwidth}
    \centering
    \includegraphics[width=0.9\textwidth]{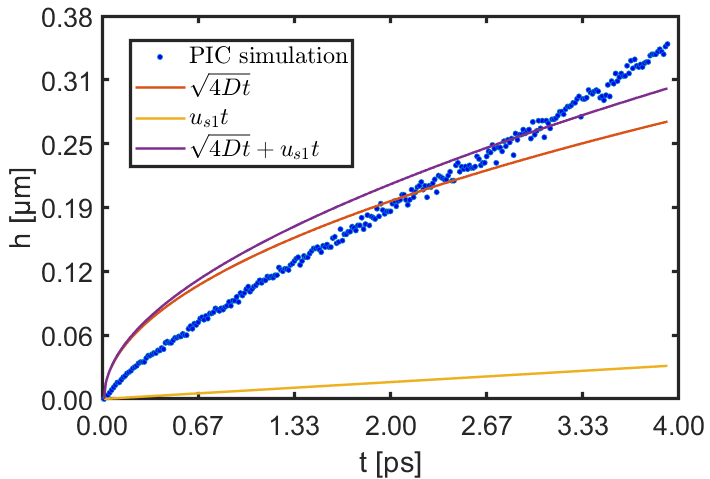}
    \caption{The diffusion height of D$^+$ ion. The blue dots is measured from the simulation, the red and orange curve corresponds to the first and second part of Eq.\ \ref{eq:z2} respectively and the purple curve is the total expression of Eq.\ \ref{eq:z2}.}
    \label{fig:Dfront}
    \end{minipage}
\end{figure}

In Fig.\ \ref{fig:Cfront} and Fig.\ \ref{fig:Dfront} we evaluated the diffusion height of C$^{6+}$ ion and D$^+$ ion in the simulations, compared with $h=\sqrt{4Dt},\ h=|u_{si}|t$ and $h=\sqrt{4Dt}+|u_{si}|t$. The diffusion height is defined by the distance between the ion fronts and the initial interface, and the ion fronts are evaluated by finding the location of a critical density of the ions, namely 1/10 of the initial ion density at the interface. As can be seen from the figures, it is evident that the mixing layer broadens in a way much different from $L_D=\sqrt{4Dt}$ under the effect of gravity. In general, Eq.\ \ref{eq:z1} and Eq.\ \ref{eq:z2} roughly fit the trend of the C$^{6+}$ and D$^+$ ion fronts respectively. The discrepancy between the simulation results and the theoretical prediction can be understood in such a way: at early times, it takes some time for the mixing layer to grow into the self-similar profile \cite{Molvig2014NonlinnearStructure}, while in our case the gravitational sedimentation at the diffusion front always makes the process slower. At late times, the mixing front will accelerate due to thermal production and expansion within the mixing layer, which is actually out of the isothermal assumption. 

We admit that Eq.\ \ref{eq:profile}, Eq.\ \ref{eq:z1} and Eq.\ \ref{eq:z2} are a very rough approximation to the exact diffusion process. However, at least in our approximation the effect of gravity on diffusion can be easily included without loss of generality, and the exponential density profile can be also taken into consideration. This allows us to carry out analyses with a very different nature from the isobaric ones.

\section{Mathematical solution and main results}
\subsection{Effect of the exponential density distribution}\label{sec:compressible}
Now that we have derived the density distribution Eq.\ \ref{eq:profile}, we are able to solve the Rayleigh equation Eq.\ \ref{eq:Rayleigh} to get the suppression factor $\Psi$. However, before we start, we would like to solve the Rayleigh equation for the exponential density profile Eq.\ \ref{eq:density} before diffusion starts, in order to measure the effect of the non-uniform density on RTI growth alone. 

In fact, analytical study about immiscible fluids with exponential distribution mentioned above have been carried out as an aspect of study of effects of compressibility on RTI \cite{Bernstein1983Effectofcompressibility,Livescu2004Compressibility}. To be specific, this effect is sometimes referred to as an static ``compressibility" effect \cite{Lafay_2007_Compressibility}, and the incompressible condition $\nabla\cdot \mathbf{u}=0$ is still applicable in this case, while a ``dynamic compressibility" refers to the effect of compressible equation of state and the adiabatic index, where the former incompressible condition has to be replaced. 

Here we stick with the static compressible effect, which is more convenient to apply to analyses. In general, the exponential density distribution has a stabilizing effect on RTI. This can be seen by solving the Rayleigh equation Eq.\ \ref{eq:Rayleigh} with the initial density distribution Eq.\ \ref{eq:density}. In either layer of plasma with $\rho_i=\rho_{i0} \text{exp}[-m_igz/((1+Z_i)k_B T)]$, the Rayleigh equation reduces to
\begin{widetext}
    \begin{equation}
    \frac{d^2 u}{dz^2}-\frac{m_i g}{(1+Z_i)k_B T}\frac{du}{dz}-k^2\left[1-\frac{m_ig^2}{(1+Z_i)k_B T\gamma^2}\right]u=0,
    \label{eq:Rayleighcomp}
\end{equation}
which has a solution of the form $u(z)=A\text{exp}(k_1 z)+B\text{exp}(k_2 z)$, where
    \begin{equation}
\begin{cases}
k_1=\frac{1}{2}\bigg[\frac{m_1 g}{(1+Z_1)k_BT}+\sqrt{\left(\frac{m_1 g}{(1+Z_1)k_BT}\right)^2+4k^2\left(1+\frac{m_1 g^2}{(1+Z_1)k_BT\gamma^2}\right)}\bigg],\\
k_2=\frac{1}{2}\bigg[\frac{m_2 g}{(1+Z_2)k_BT}-\sqrt{\left(\frac{m_2 g}{(1+Z_2)k_BT}\right)^2+4k^2\left(1+\frac{m_2 g^2}{(1+Z_2)k_BT\gamma^2}\right)}\bigg],
\end{cases}
\label{eq:solutioncomp}
\end{equation}
\end{widetext}
which is actually the same as the result of Berstein \cite{Bernstein1983Effectofcompressibility}, where the sound velocity is $c_i=\sqrt{(1+Z_i)k_BT/m_i}$ in our case. With the boundary condition at infinity, we can deduce $u(z)=A\text{exp}(k_2 z)$ for $z>0$ and $A\text{exp}(k_1 z)$ for $z<0$. It is not hard to see that $k_1>k$ and $k_2<-k$. As a result, when we integrate Eq.\ \ref{eq:Rayleigh} across the interface,
\begin{equation}
    \rho_2 k_2-\rho_1 k_1=k^2 \frac{g}{\gamma^2}(\rho_1-\rho_2),
\end{equation}
we get the implicit dispersion relation of $\gamma$
\begin{equation}
    \gamma^2=-gk^2\frac{\rho_2-\rho_1}{\rho_2k_2-\rho_1k_1}<gk\frac{\rho_2-\rho_1}{\rho_2+\rho_1}.
    \label{eq:compressible}
\end{equation}
This difference will become important when the characteristic length $L_g=(1+Z_i)k_BT/m_ig$ is evidently smaller than $\lambda/4\pi$. In previous simulation settings, the former length is 0.44 $\mu$m for the upper plasma of C$^{6+}$ and 2.28 $\mu$m for the lower plasma of D$^+$, so this effect will matter when the perturbation wavelength $\lambda>5.53\ \mu $m.

Besides, the dispersion relation Eq.\ \ref{eq:compressible} has been solved analytically \cite{Bernstein1983Effectofcompressibility}:
\begin{equation}
    \frac{\gamma^2}{kg}=\frac{\sqrt{1+G^2}-1}{S}
\end{equation}
Where $G=k[(1+Z_1)/m_1-(1+Z_2)/m_2]k_BT/g$ and $S=k[(1+Z_1)/m_1+(1+Z_2)/m_2]k_BT/g$. It is clear that the growth relation is only related to $Z_i,m_i,T$ and $g$.

\begin{figure}[h]
\centering
\includegraphics[width=0.9 \columnwidth]{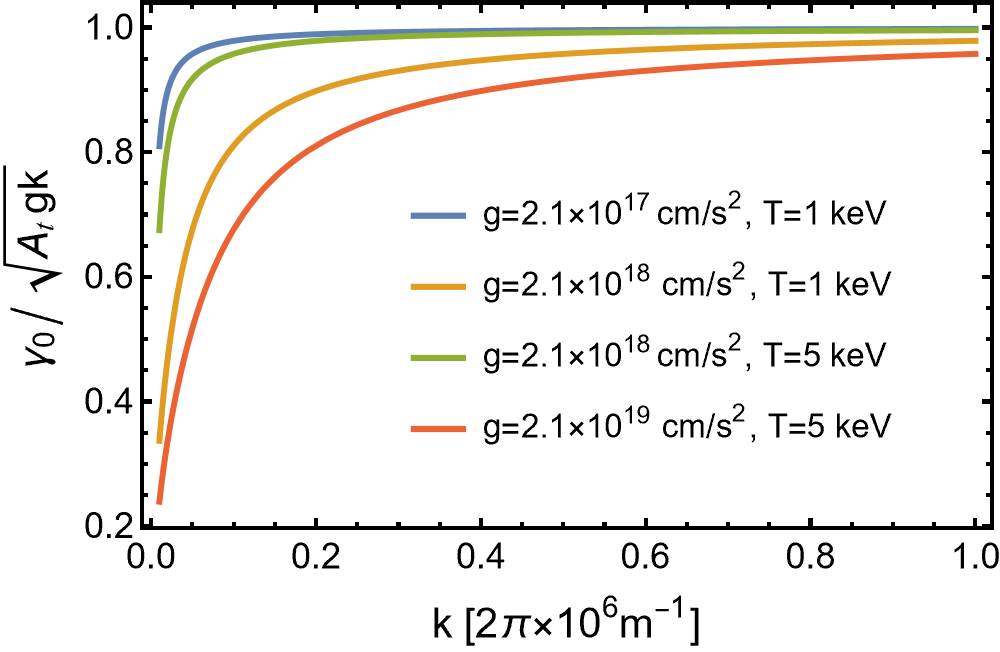}
\caption{Dispersion relation from the solution of Eq.\ \ref{eq:compressible}. The parameters $m_i$ and $Z_i$ are determined as described in Sec.\ \ref{sec:model}.}
\label{fig:comp}
\end{figure}
Fig.\ \ref{fig:comp} shows the dispersion relation of the growth rate $\gamma$ containing the static compressible effect. As can be seen, $\gamma$ tends to 0 for small k and approaches unity asymptotically as k gets larger. Larger gravity $g$ will result in a larger damping effect and higher temperature $T$ will simply reduces the effect. For the $Z_i,m_i$ set we are concerned about, the growth rate $\gamma$ only dampens obviously when the ratio of acceleration to temperature is large, as shown by the red curve, where $g=2.1\times10^{19}\ \text{cm/s}^2,T=5\ \text{keV}$, corresponding to the simulation settings \cite{Yin2016Plasmakineticeffectsoninterfacialmix}. For RTI in the ICF deceleration stage, the acceleration is about 100 times smaller although the temperature is 5 times lower, and the effect only becomes important when the wavenumber is very small or the perturbation wavelength is longer than about 50$\mu $m, as shown by the blue curve.

\subsection{Time-dependent solution of the equation}
Now that we have derived the time-dependent density profile Eq.\ \ref{eq:profile}, we can solve the Rayleigh equation Eq.\ \ref{eq:Rayleigh} in order to get the corresponding time-dependent growth rate $\gamma(t)$. As the density profile is piecewise, obviously the eigenfunction of the Rayleigh equation is also piecewise. For the unmixed region $z<z_1(t)$ and $z>z_2(t)$, the eigenfunction has been derived already in Sec.\ \ref{sec:compressible}. For the mixing layer $z_1(t)<z<z_2(t)$, the density is linear and the eigenfunction $u(z)$ can be expressed as a combination of confluent hypergeometric functions \cite{cherfils2000analytic}. To sum up, the overall eigenfunction is
\begin{widetext}
    \begin{equation}
    u(z)=\begin{cases}
    C_1 e^{k_1 z},& \text{if } z < z_1(t), \\
    B_1 e^{-kz} M(a,1,x(z))+B_2 e^{-kz} U(a,1,x(z)),& \text{if } z_1(t)\leq z \leq z_2(t),\\
    C_2 e^{k_2 z},& \text{if } z > z_2(t),
    \end{cases}
    \label{eq:eigenfunction}
\end{equation}
where $M(a,1,x)$ and $U(a,1,x)$ are the two different types of confluent hypergeometric function, and they are linear independent unless $a$ is an integer. $a(\gamma)=\frac{1}{2}(1-kg/\gamma^2)$ is a non-dimensional function of $\gamma$ and as $\gamma<\sqrt{A_tgk}$ always holds, the range of $a$ is $a<(1-1/A_t)/2<0$. $x(z)=2kz-2k(\rho_1 z_2-\rho_2 z_1)/(\rho_1-\rho_2)$ is a non-dimensional function of $z$. $C_1,C_2,B_1,B_2$ are coefficients which need to be determined by continuous conditions. As the density profile is continuous, both $u(z)$ and its derivative $\partial u(z)/\partial z$ should be continuous throughout the area, and making use of the continuous condition at $z_1(t)$ and $z_2(t)$ will give us the relations between $C_1,C_2,B_1,B_2$ and the growth rate $\gamma$. The continuous condition at the turning points $z_1$ and $z_2$ results in four equations:
    \begin{align}
C_1&=e^{-(k+k_1)z_1}[B_1 M(a,1,x_1)+B_2 U(a,1,x_1)],\\
(k_1+k)C_1&=e^{-(k+k_1)z_1}2k[B_1 M'(a,1,x_1)+B_2 U'(a,1,x_1)],\\
C_2&=e^{-(k+k_2)z_2}[B_1 M(a,1,x_2)+B_2 U(a,1,x_2)],\\
(k_2+k)C_2&=e^{-(k+k_2)z_2}2k[B_1 M'(a,1,x_2)+B_2 U'(a,1,x_2)],
\end{align}
\end{widetext}

where $x_i=x(z_i)$, and $M'(a,1,x)$, $U'(a,1,x)$ mark $\partial M(a,1,x)/\partial x$, $\partial U(a,1,x)/\partial x$ respectively. To find a non-trivial solution of $u(z)$, the determinant of $C_1,C_2,B_1,B_2$ must be zero. Note that $k_1,k_2$ are both functions of $\gamma$, the total determinant can be transformed into a function Det$(a)$ of the non-dimensional variable $a(\gamma)$, and the zeros of Det$(a)$ corresponds to an eigenvalue $\gamma$ and the corresponding eigenfunction $u(z)$, where the set of coefficients is determined. Different from the classical uniform case, Det$(a)=0$ has an infinite set of solutions \cite{cherfils2000analytic}, which means different eigenmodes of $u(z)$ can occur for a single perturbation.

\begin{figure}[h]
\centering
  \begin{minipage}[t]{0.49\textwidth}
    \centering
    \includegraphics[width=0.9\textwidth]{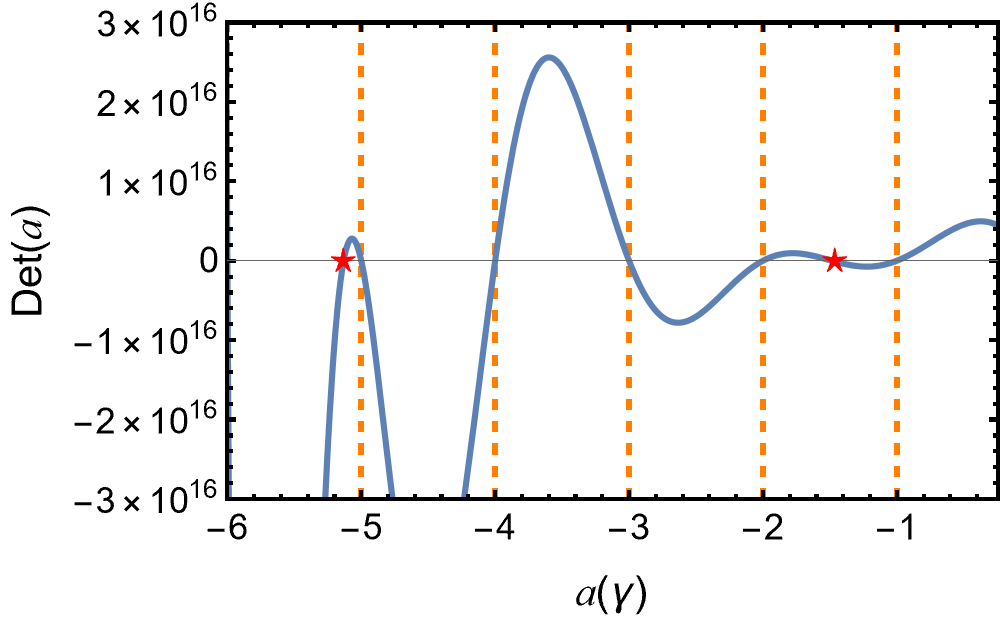}
    \caption{The determinant function Det$(a)$ for a perturbation of a wavelength of $1.28\mu $m at t=2 ps.}
    \label{fig:determinant}
  \end{minipage}
  \begin{minipage}[t]{0.49\textwidth}
    \centering
    \includegraphics[width=0.9\textwidth]{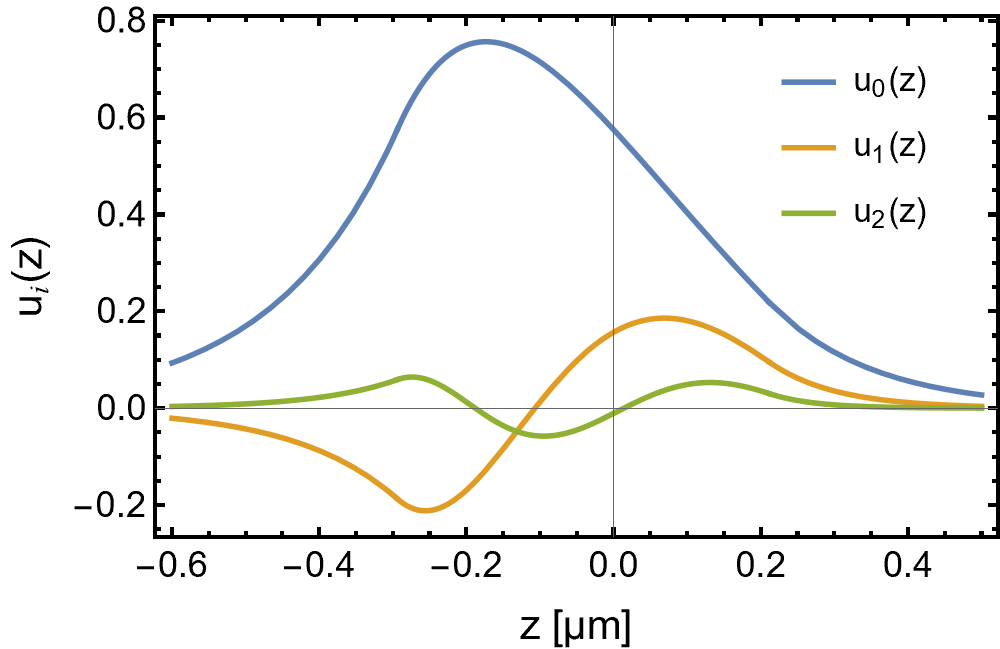}
    \caption{The eigenfunctions $u_i(z)$ of three fastest modes, normalized by setting the coefficient $C_2=1$.}
    \label{fig:eigenfunction}
  \end{minipage}
\end{figure}

It is not difficult to find the zeros of Det$(a)$ numerically. Fig.\ \ref{fig:determinant} shows the Det$(a)$ of a perturbation $\lambda=1.28\ \mu$m at $t$=2 ps. Note that $M(a,1,x)$ and $U(a,1,x)$ are linear dependent when $a$ is an integer, non-trivial solutions only exist for the non-integer zeros. The first non-integer zero is at $a_0=-1.46$ and the second zero is at $a_1=-5.13$. A third zero is at $a_2=-12.5$ which is not shown in the figure. The corresponding growth rate can be recovered by $\gamma=\sqrt{gk/(1-2a)}$, and $\gamma_0=0.61\sqrt{A_tgk}, \gamma_1=0.36\sqrt{A_tgk}, \gamma_2=0.24\sqrt{A_tgk}$. In Fig.\ \ref{fig:eigenfunction} the corresponding eigenfunctions $u_0(z),u_1(z)$ and $u_2(z)$ are plotted. It can be seen that the zeros of the eigenfunction accords with the order of the zeros, which is a direct result of Liouville theorem. 

In RTI, an arbitrary initial perturbation can be decomposed into combinations of different eigenfunctions, and they will compete with each other. However, the first eigenmode $u_0(z)$ has the largest growth rate $\gamma_0$ among all the eigenmodes, and is expected to dominate the instability growth. As a result, we will focus on the largest growth rate $\gamma_0$ for different perturbation wavelength.

\subsection{Main results and discussion}

\begin{figure}[h]
\centering
  \begin{minipage}[t]{0.49\textwidth}
    \centering
    \includegraphics[width=0.9\textwidth]{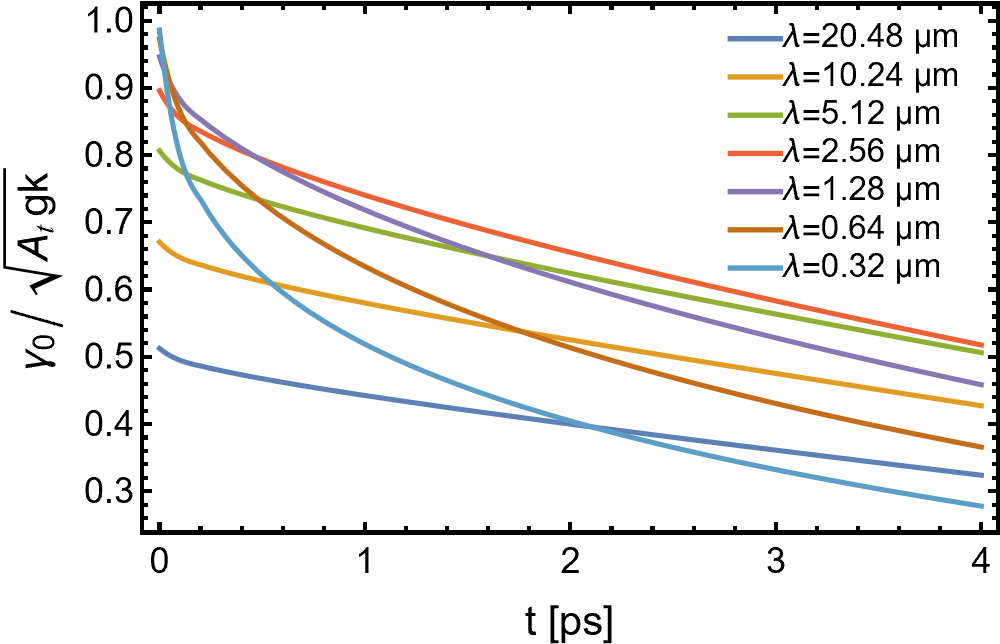}
    \caption{Normalized growth rate $\gamma_0/\sqrt{A_tgk}$ for different perturbation wavelengths $\lambda$.}
    \label{fig:growth1}
  \end{minipage}
  \begin{minipage}[t]{0.49\textwidth}
    \centering
    \includegraphics[width=0.9\textwidth]{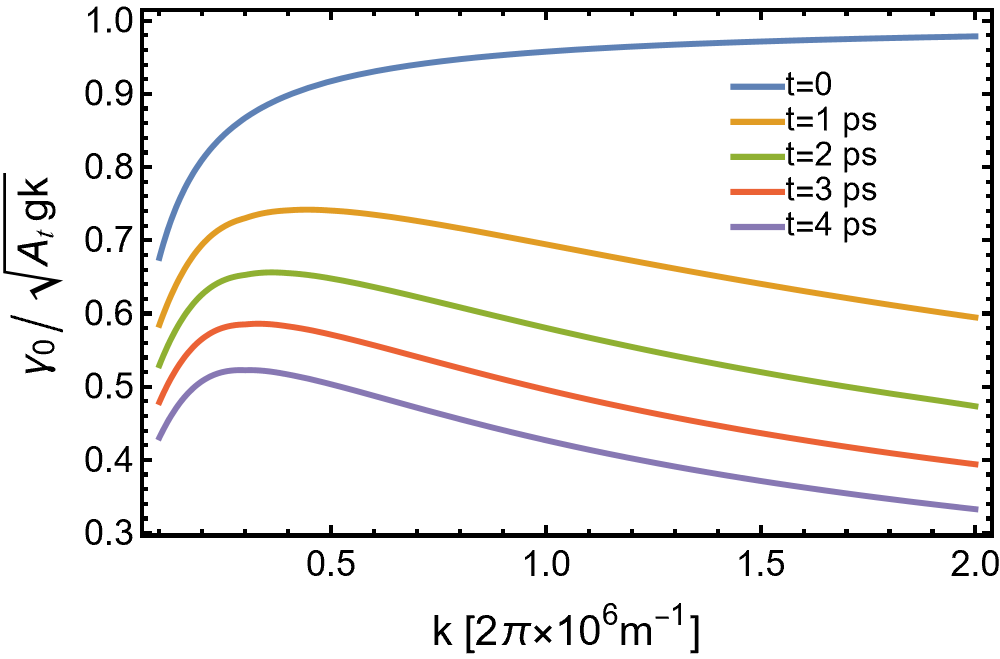}
    \caption{Normalized growth rate $\gamma_0/\sqrt{A_tgk}$ for different diffusion timescales $t$.}
    \label{fig:growth2}
  \end{minipage}

\end{figure}

We then solved the largest growth rate $\gamma_0$ for different perturbation wavelength and diffusion time, with the set of parameters introduced in Sec.\ \ref{sec:model}. Fig.\ \ref{fig:growth1} shows the evolution of growth rate for different perturbation wavelengths. Although the growth rate of all wavelengths decreases with time as the mixing layer broadens, how strongly the growth rate of different wavelengths is dependent on time is clearly distinct. In the beginning, the exponential density profile or compressible effect dominates the damping of the growth rate, where perturbations of larger wavelength or smaller wavenumber are more affected, as predicted in Fig.\ \ref{fig:comp}. However, as the mixing layer starts to broaden, perturbations of smaller wavelength or larger wavenumber seem to be more sensitive to the span of the mixing layer, and the growth rate damps with time much faster than perturbations with larger wavelength. As a result, the growth rate is a combination of two effects and will not exhibit a monotonicity over perturbation wavelength unless in the very beginning.

Fig.\ \ref{fig:growth2} shows the dispersion relation of growth rate at different diffusion timescales. Except for $t=0$ when the mixing layer is not formed yet, the non-monotonicity of growth rate over the perturbation wavenumber $k$ is clearly shown in the figure. The non-monotonicity can be understood as follows: perturbation of smaller wavenumber is dominated by the damping effect of the exponential density distribution outside the mixing layer, while the growth rate of larger perturbation wavenumber is mainly damped by the existence of the mixing layer. As a result, a maximum of growth rate occurs for an intermediate wavenumber. As the mixing layer broadens with time, the damping of larger wavenumber becomes stronger. As a result, the maximum wavenumber of growth rate tends to move to smaller wavenumber with time, as can be seen in the figure.

\begin{figure}[htbp]
    \centering
    \includegraphics[width=0.9\linewidth]{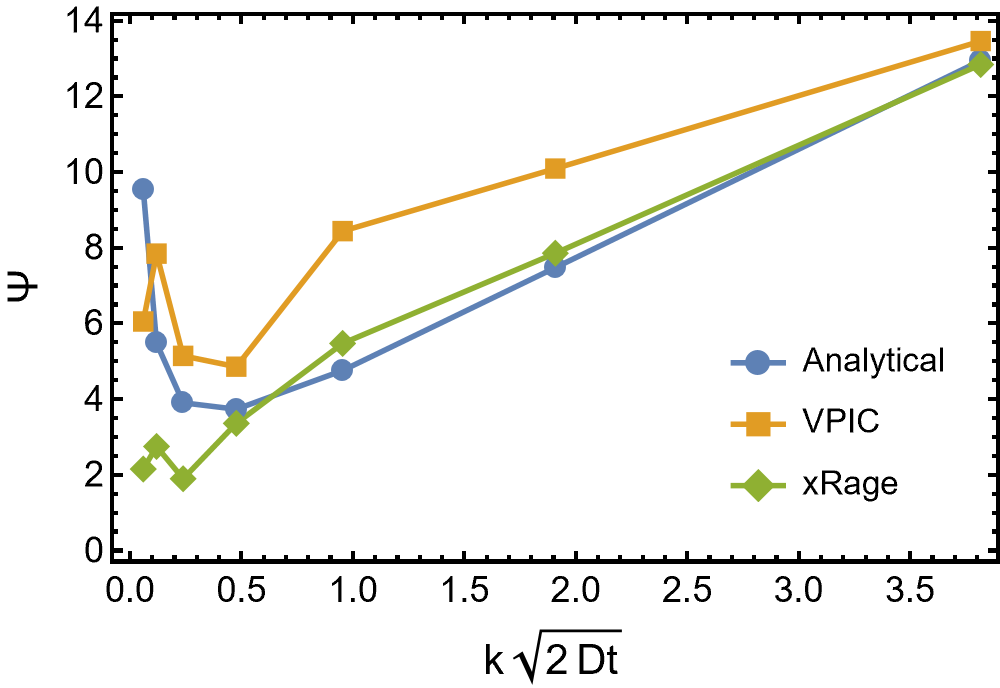}
    \caption{Density suppression factor $\Psi$ of Keenan's style, where $\Psi=A_tgk/\gamma_0^2$. The yellow dots and green dots are subtracted from the VPIC and xRage simulations carried out in \cite{Keenan2023ImprovedAnalytica}.}
    \label{fig:factor}
\end{figure}

We might also compare our result with Keenan's solution of the density suppression factor $\Psi$ \cite{Keenan2023ImprovedAnalytica}. The yellow curve shows our solution of the density suppression factor $\Psi$ evaluated at $t=4\ \text{ps}$. Compared with Fig. 7 in  \cite{Keenan2023ImprovedAnalytica}, our result is closer to the $\Psi$ inferred from simulations, which shows non-monotonicity over wavelengths. Our result is much closer to the VPIC results at small wavenumbers. For large wavenumbers, our result is still smaller than the VPIC results, which indicates that our treatment of the diffusion effect is still not exact, as some rough approximations are made to describe the density distribution of the mixing layer. Different from our linear density distribution, the realistic density distribution of the mixing layer should be smoother at both ends, which leads to larger damping of the instability, as observed in RMI simulations \cite{sano2020suppression}. We also note that we should actually adopt boundary conditions at the boundaries of the simulation domain to solve Eq.\ \ref{eq:Rayleighcomp}, while we adopted conditions at infinity instead. However, direct calculation shows that the difference is negligible, as the simulation domain is not so small compared to the perturbation wavelength.

Moreover, our treatment is not able to explain the other non-monotonicity at extremely small wavenumbers as exhibited in the VPIC and xRage simulations. It might result from the incompressible nature of our analysis, as we are still adopting $\nabla\cdot \mathbf{u}=0$. A full inclusion of compressibility may help improve upon this, while it may be difficult as the compressibility of the plasma  may be very different across the interface.

\section{Conclusion}

In this paper, we roughly modelled the effect of mass diffusion on RTI under a large gravity. We point out that a large gravity will lead to an exponential density distribution under isothermal hydrodynamic equilibrium, and will also accelerate the binary diffusion process by floating up of light ions and sedimentation of heavy ions, which was often neglected in former diffusive RTI researches. Then we modelled the time-dependent density profile of the binary plasma with a linear density approximation to the mixing layer. We admit that the approximation is rough, but it manages to illustrate the decrease of the effective Atwood number under the exponential density distribution, and also included the effect of gravity-accelerated diffusion by adding the sedimentation displacement to the location of the boundaries of the mixing layer. Finally, we solve the Rayleigh equation with the time-dependent density distribution and derive the time-dependent growth rate $\gamma_0$. We find out that the density suppression factor $\Psi=A_tgk/\gamma_0^2$ is not monotonic over the perturbation wavenumber $k$, as perturbations of small wavenumber are damped by the exponential density and perturbations of large wavenumbers are damped due to the formation of the mixing layer. This result reaches better agreement with former simulations\ \cite{Keenan2023ImprovedAnalytica, Yin2016Plasmakineticeffectsoninterfacialmix, Vold2021Plasmatransport}.

We should note that our modelling of the mixing layer is far from accurate, and we expect an analytical solution of this problem, which of course is difficult. So far, the analytical descriptions of plasma binary diffusion are mostly based on assumptions like isothermal and isobaric conditions, which can simplify the problem very much. However, under a large gravity the isobaric condition will fail, as the hydrodynamic equilibrium requires a pressure gradient against the gravity. Also, a realistic diffusion process must be accompanied by heating of the mixing layer, which will violate the isothermal condition.

Still, although we take the effect of gravity into consideration, our result is not strictly consistent with the simulation results. On one hand, this may result from our incomplete analyses, as we are still adopting incompressible equations, and the diffusion coefficient is considered a constant while it actually varies with density and temperature. On the other hand, the initial perturbations in simulations are often not true eigenmodes of the Rayleigh equation, so the growth rate will naturally differ from the growth rate calculated for the eigenmodes.

After all, the effects of many physical factors on RTI are still ambiguous even after over 140 years' research, while the theories are becoming more and more complicated. We still hope for some breakthroughs leading to more integrated understanding of RTI, which will greatly assist us with the improvement of ICF designs.

This work was supported by the Strategic Priority Research Program of Chinese Academy of Sciences (Grant No. XDA250010100 and XDA250050500), National Natural Science Foundation of China (Grant No. 12075204) and Shanghai Municipal Science and Technology Key Project (Grant No. 22JC1401500). D. Wu thanks the sponsorship from Yangyang Development Fund.





\bibliography{Reference}

\end{document}